\documentclass[11pt]{article}
\usepackage{amsmath, amssymb}
\usepackage{geometry}
\usepackage{graphicx}
\usepackage{setspace}
\usepackage{bm}
\usepackage{mathrsfs}
\usepackage{authblk}
\usepackage{enumitem}
\usepackage{csquotes}
\usepackage[
  style=apa,
  backend=biber,
  maxnames=2,
  minnames=1,
  date=year 
]{biblatex}
\DeclareLanguageMapping{english}{american-apa}
\usepackage{xpatch}

\DeclareCiteCommand{\textcite}
  {}
  {\printtext[bibhyperref]{%
     \printnames{labelname}~(\printdate)}
  }
  {\multicitedelim}
  {}

\DeclareCiteCommand{\parencite}
  {\bibopenparen}
  {\bibhyperref{\printnames{labelname},\addspace\printdate}}
  {\multicitedelim}
  {\bibcloseparen}

\usepackage[dvipsnames]{xcolor}
\usepackage[
    colorlinks=true,
    linkcolor=blue,      
    citecolor=OrangeRed,   
    urlcolor=teal        
]{hyperref}

\title{\textbf{AI Agents and the Attention Lemons Problem in Two-Sided Ad Markets}}
\author{
  Md Mahadi Hasan\textsuperscript{\dag} \\
  \textit{North Dakota State University}
}

\date{July 31, 2025}
\begin{document}

\maketitle
\begingroup
  \renewcommand\thefootnote{\dag}%
  \footnotetext{%
    \noindent
    \begin{minipage}[t]{\linewidth}
      \raggedright                
      Department of Agribusiness and Applied Economics, North Dakota State University.\\[0.5ex]
      Email: \texttt{mdmahadi.hasan.1@ndsu.edu}\\[0.5ex]
      ORCID: \url{https://orcid.org/0009-0009-5728-5119}
    \end{minipage}%
  }%
\endgroup
\begin{abstract}
I develop a theoretical model to examine how the rise of autonomous AI (artificial intelligence) agents disrupts two-sided digital advertising markets. Through this framework, I demonstrate that users’ rational, private decisions to delegate browsing to agents create a negative externality, precipitating declines in ad prices, publisher revenues, and overall market efficiency. The model identifies the conditions under which publisher interventions such as blocking AI agents or imposing tolls may mitigate these effects, although they risk fragmenting access and value. I formalize the resulting inefficiency as an “attention lemons” problem, where synthetic agent traffic dilutes the quality of attention sold to advertisers, generating adverse selection. To address this, I propose a Pigouvian correction mechanism: a per-delegation fee designed to internalize the externality and restore welfare. The model demonstrates that, for an individual publisher, charging AI agents toll fees for access strictly dominates both the `Blocking' and `Null (inaction)' strategies. Finally, I characterize a critical tipping point beyond which unchecked delegation  triggers a collapse of the ad-funded digital market.
\end{abstract}

\vspace{4em}
\noindent\textbf{Keywords:} \textit{AI agents; attention economy; adverse selection; externalities;  two-sided ad markets} \\

\vspace{0.3em}
\noindent\textbf{JEL Codes:} \textit{L13, L86, M37, D62, D82}

\clearpage
\section{Introduction}
The business model of major digital platforms, most notably Google, is built on a foundational exchange: users receive free content and services, and advertisers pay to access human attention through search and display ads. This implicit contract forms the basis of the monetization of digital media, with search engines, publishers, and platforms deriving revenue from ad impressions and clicks, which are presumed to be made by real people with real purchasing intent. But this equilibrium is now under quiet disruption  with the rise of autonomous AI (artificial intelligence) agents.

AI agents, by their very design, do not behave like humans. They do not browse with emotions, preferences, or purchasing intent. As a result, they neither click on advertisements nor respond to visual marketing cues. Their interaction with web content is typically utilitarian and task-driven, focused on extracting information rather than engaging with the surrounding environment. This fundamental behavioral gap means that conventional ad delivery systems, which rely on capturing human attention and influencing decision-making, are ineffective when the end user is an autonomous agent. The presence of such agents introduces a new class of web traffic that consumes content without contributing to the monetization ecosystem.

Advertisers, who believe they are paying for human attention, are increasingly paying to influence bots, a reality that renders the core assumptions of digital marketing obsolete and signals the beginning of a profound shift. This transformation introduces a fundamental market inefficiency, conceptualized as an ``attention lemons" problem. Advertisers, unable to distinguish between genuine human attention and valueless synthetic agent interactions, face uncertainty and adverse selection problem, analogous to the market failure described by \textcite{akerlof1970lemons} in his seminal work. In this new context, AI agent-generated attention acts as the low-quality ``lemon," while human attention is the high-quality ``peach." As the share of AI traffic grows, the average quality of attention in the market declines, depressing market-clearing prices for ad impressions and eroding publisher revenues. In addition, inability to distinguish quality leads to a pooling mispricing.

In this paper, I develop a theoretical model to explore the economic implications of AI-mediated browsing behavior. I characterize the emergent inefficiencies and market failures precipitated by AI agent delegation, examine publisher strategies for adaptation, propose mechanisms for correcting these distortions and identify a critical threshold for the collapse of the ad-funded market.

The analysis is grounded in the literature on two-sided markets, which studies platforms that act as intermediaries between two distinct groups of users, managing the cross-side network externalities that exist between them. Foundational works by \textcite{rochet2003platform} and \textcite{armstrong2006competition} established the principles of platform competition. A key insight from this literature is the critical distinction between the platform's price level (the total fee extracted from a transaction) and its price structure (the allocation of that fee between the two sides). Platforms often subsidize the more price-sensitive side to ``get both sides on board," thereby increasing participation and maximizing the value of the network externality. 

Advertiser-supported media is a classic example of a two-sided market, a structure analyzed in foundational work by \textcite{rysman2009economics}.  In this model, platforms like newspapers or digital publishers often heavily subsidize or provide free content to one side (readers/users) to build a large audience, which in turn makes the platform more attractive to the other side (advertisers). The model presented in this paper adopts this precise structure: publishers serve as platforms mediating between a user base and a pool of advertisers. The price of an ad impression, \textit{p(a)}, is endogenously determined by the quality and composition of the user base, a direct application of two-sided pricing theory. This paper builds upon this established work by introducing a novel technological shock, AI agent delegation, which endogenously alters the composition and, therefore, the value of one side of the market (the user base). This directly degrades the positive cross-side externality that advertisers pay to access, creating a new dynamic not previously explored in the platform literature.

The conceptualization of the market rests on the idea that human attention is a scarce and valuable economic resource. This notion was first articulated by \textcite{simon1996designing}, who observed that ``a wealth of information creates a poverty of attention," thereby framing the allocation of attention as a fundamental economic problem. More recently, \textcite{wu2017attention}  provided a historical account of how entire industries, \textit{the attention merchants}, have evolved to systematically harvest, bundle, and resell this scarce human attention. The modern ad-supported internet represents the evolved form of this business model, with sophisticated platforms designed to maximize user engagement and capture moments of attention that can be sold to advertisers. This paper operationalizes these concepts within its formal model.
The advertiser's valuation, \textit{v}, represents the economic price for a unit of scarce, monetizable human attention. The rise of AI agents, which can process information at a massive scale without the cognitive costs or scarcity constraints of humans, represents a paradigm shift in the attention economy. The \textit{attention lemons} problem arises precisely because these agents provide a substitute for information processing but not for the scarce, commercially valuable attention that underpins the market's viability.

This paper distinguishes itself from adjacent literature in several key ways. First, it departs from existing research on non-human traffic (NHT) and ad fraud. That body of work, exemplified by economic models like \textcite{mungamuru2008should} and detection systems like \textcite{Pastor2019Nameles}, typically treats bot traffic as an exogenous and malicious phenomenon to be filtered out. In those models, invalid traffic is a parameter, not an equilibrium outcome. In contrast, this paper models the share of AI traffic, \textit{a}, as the result of endogenous, utility-maximizing decisions by rational users. 

Second, the publisher adaptation strategies analyzed here connect to and extend the game-theoretic literature on ad-blocking. The \textit{blocking} strategy is a direct analogue to publishers deploying ad-block walls to force users to view ads. The \textit{tolling} strategy, however, introduces a novel mechanism. It differs significantly from the whitelisting negotiations modeled by \textcite{gritckevich2022ad}, in which ad-blockers extract payments from publishers. Here, the revenue flow is reversed: publishers directly charge AI agents a fee for API (application programming interface) access. This represents a plausible business-to-business (B2B) revenue model for content in an AI  agent–mediated advertising market.

\clearpage
The remainder of the paper is structured as follows. Section 2 presents the model. Section 3 outlines the attention lemons market failure. Section 4 analyzes publisher strategies. Section 5 introduces a corrective mechanism. Section 6 examines the tipping point for market collapse. Section 7 concludes.

\section{Model}

The model considers a digital ecosystem comprising four classes of economic agents: \textit{publishers}, \textit{human users}, \textit{AI agents}, and \textit{advertisers}. $N$ is a set of publishers, indexed by $i = 1, \ldots, N$.  I model a market of N publishers operating under \textit{ monopolistic competition}. Each publisher $i$ offers \textit{differentiated content}, making it an \textit{imperfect substitute} for other publishers. This differentiation grants each publisher a degree of market power over its specific content.

Total web traffic is normalized to 1. A fraction $a \in [0, 1]$ of this traffic is generated by autonomous AI agents. The remaining fraction, $1 - a$, is generated by humans browsing directly. Advertisers value an ad impression viewed by a human at $v > 0$. They value an impression "viewed" by an AI agent at 0, as agents do not have purchase intent. Advertisers can observe the aggregate share of AI traffic, $a$, but cannot distinguish the source (human or AI) of any individual impression. \footnote{Throughout the paper, AI traffic and AI agent--generated traffic are used interchangeably.} In a competitive advertising market, the price of an impression, $p(a)$, will equal its expected value:

\[
p(a) = \mathbb{E}[\text{Value}] = v \cdot \mathbb{P}(\text{Human}) + 0 \cdot \mathbb{P}(\text{AI})
\]

\begin{equation}
p(a) = v(1 - a) + 0 \cdot a = v(1 - a)
\label{eq:1}
\end{equation}

This is the expected price per impression in the market. Each publisher $i$ receives a traffic share $s_i \geq 0$, with the shares summing to one:
\[
\sum_{i=1}^{N} s_i = 1
\]
The revenue for publisher $i$ is given by:

\[
R_i(a) = s_i \cdot p(a) = s_i v(1 - a)
\]

Each human user chooses to delegate browsing to an AI agent if the net utility gain from doing so is positive. The utility gain is $\Delta u > 0$, and each user has an idiosyncratic, private cost of delegation $c$, drawn from a distribution with a strictly positive density function $f(c)$ and a cumulative distribution function $F(c)$. A user delegates if $\Delta u - c > 0$. The aggregate share of users who delegate determines the AI traffic share, $a$. Thus, the privately chosen equilibrium level of delegation is:

\begin{equation}
a^{\text{priv}} = \mathbb{P}(c < \Delta u) = F(\Delta u)
\label{eq:2}
\end{equation}

\section{The Attention Lemons Market Failure} \label{sec:lemons_failure}

 I now establish the existence of a market failure resulting from AI agent-generated traffic. The key result is formalized in ~\hyperlink{thm:AI_market_failure}{Theorem 1}. Before presenting the theorem, I first prove that the social welfare function is strictly concave in the share of AI traffic.\\

\noindent
\textbf{\hypertarget{lem:welfare_concavity}{LEMMA 1 (Welfare Concavity)}} \\
\textit{The social welfare function $W(a)$ is strictly concave in the share of AI traffic $a$ if the cost density function is strictly positive, $f(c) > 0$, and the price function is weakly concave, $\frac{\partial^2 p}{\partial a^2} \leq 0$.}

\medskip
\noindent\textbf{Proof.} \\
The social welfare function is:

\begin{equation}
W(a) = 
\underbrace{
\int_0^{F^{-1}(a)} (\Delta u - c) f(c) \, dc
}_{\text{User Surplus}}
+
\underbrace{p(a)}_{\text{Total Publisher Revenue}}
+
\underbrace{am}_{\text{AI Agent Profit}}
\label{eq:3}
\end{equation}

where $m \geq 0$ is the exogenous profit an AI provider earns per delegation.

To determine its concavity, I compute its second derivative with respect to $a$. First, I find the first derivative, $\frac{\partial W}{\partial a}$, by applying the Leibniz rule to the integral term:
\[
\frac{\partial W}{\partial a} = \frac{d}{da} \left[ \int_0^{F^{-1}(a)} (\Delta u - c) f(c) \, dc \right] + \frac{\partial p}{\partial a} + m
\]

The derivative of the integral term is $(\Delta u - F^{-1}(a)) f(F^{-1}(a)) \cdot \frac{d}{da} \left[ F^{-1}(a) \right]$. By the definition of an inverse function, we have $F(F^{-1}(a)) = a$. Differentiating both sides with respect to $a$ gives:
\[
f(F^{-1}(a)) \cdot \frac{d}{da} \left[ F^{-1}(a) \right] = 1
\]

This implies that:
\[
\frac{d}{da} \left[ F^{-1}(a) \right] = \frac{1}{f(F^{-1}(a))}, \quad \text{assuming } f(\cdot) > 0
\]

Substituting this result back into the derivative of the integral gives:
\[
(\Delta u - F^{-1}(a)) f(F^{-1}(a)) \cdot \left( \frac{1}{f(F^{-1}(a))} \right) = \Delta u - F^{-1}(a)
\]

Thus, the first derivative of the social welfare function is:
\begin{equation}
\frac{\partial W}{\partial a} = (\Delta u - F^{-1}(a)) + \frac{\partial p}{\partial a} + m
\label{eq:4}
\end{equation}

Differentiate a second time with respect to $a$:
\begin{equation}
\frac{\partial^2 W}{\partial a^2} = \frac{d}{da} \left[ \Delta u - F^{-1}(a) \right] + \frac{\partial^2 p}{\partial a^2} + \frac{d}{da}[m]
\label{eq:5}
\end{equation}

\[
\frac{\partial^2 W}{\partial a^2} = - \frac{d}{da} \left[ F^{-1}(a) \right] + \frac{\partial^2 p}{\partial a^2}
\]

Substituting the expression for $\frac{d}{da} \left[ F^{-1}(a) \right]$ again:
\[
\frac{\partial^2 W}{\partial a^2} = - \frac{1}{f(F^{-1}(a))} + \frac{\partial^2 p}{\partial a^2}
\]

For $W(a)$ to be strictly concave, we require $\frac{\partial^2 W}{\partial a^2} < 0$. The lemma provides two conditions. First, $f(c) > 0$ for all $c$ in the support. This ensures that the term $-\frac{1}{f(F^{-1}(a))}$ is well-defined and strictly negative. Second, $\frac{\partial^2 p}{\partial a^2} \leq 0$. This ensures that the second term is non-positive. The sum of a strictly negative term and a non-positive term is strictly negative. Therefore, under the given conditions, $\frac{\partial^2 W}{\partial a^2} < 0$, which proves that the social welfare function $W(a)$ is strictly concave. The model’s specific price function, $p(a) = v(1 - a)$, satisfies this condition as its second derivative is zero.  \quad$\blacksquare$
 
\bigskip
\noindent
\textbf{\hypertarget{thm:AI_market_failure}{THEOREM 1:}} \textit{For any positive advertiser valuation ($v > 0$), the presence of AI agent traffic ($a > 0$) induces a market failure characterized by: Publisher revenue strictly decreases as the share of AI traffic increases. The privately chosen level of delegation is excessive, compared to the social optimum, creating a deadweight loss. The market systematically misprices individual impressions, with advertisers overpaying for AI-generated impressions and underpaying for human-generated impressions, leading to the aforementioned inefficiency.}
\bigskip

\noindent\textbf{Proof.} 
\textit{Part 1 (Strict Revenue Decay)}  Publisher $i$’s revenue is $R_i(a) = s_i v (1 - a)$. To determine the effect of an increase in AI traffic, I differentiate $R_i(a)$ with respect to $a$:

\[
\frac{dR_i}{da} = \frac{d}{da} \left[ s_i v (1 - a) \right] = -s_i v
\]

Since $s_i > 0$ and $v > 0$ by assumption, it follows that:

\[
\frac{dR_i}{da} < 0 \quad \forall i
\]

This proves that as the share of non-monetizable AI traffic $a$ rises, the revenue for every publisher strictly declines.

\textit{Part 2 (Deadweight Loss)}  
To assess efficiency, I compare the private equilibrium $a^{priv}$ with the socially optimal level $a^*$ that maximizes total social welfare, $W(a)$. Social welfare is the sum of user surplus, total publisher revenue, and AI agent provider profits.

The social welfare function is given by:
\[
W(a) = \int_0^{F^{-1}(a)} (\Delta u - c) f(c) \, dc + p(a) + am
\] 

Total publisher revenue is $p(a)$. This is because total publisher revenue is the sum of individual publisher revenues. Since the price $p(a)$ is common to all publishers, this sum is:

\[
\sum_{i=1}^{N} R_i(a) = \sum_{i=1}^{N} s_i p(a) = p(a) \sum_{i=1}^{N} s_i = p(a) \cdot 1 = p(a)
\]

By \hyperlink{lem:welfare_concavity}{Lemma 1}, $W(a)$ is strictly concave. Now, I find the socially optimal level $a^*$ by solving the first-order condition $\frac{dW}{da} = 0$. Using the Leibniz rule on the integral term:

\[
\frac{d}{da} \int_0^{F^{-1}(a)} (\Delta u - c) f(c) \, dc = (\Delta u - F^{-1}(a)) \cdot f(F^{-1}(a)) \cdot \frac{d}{da} F^{-1}(a)
\]

By the definition of the inverse function, $a = F(F^{-1}(a))$. Differentiating with respect to $a$ gives:

\[
1 = f(F^{-1}(a)) \cdot \frac{d}{da} F^{-1}(a) \quad \Rightarrow \quad \frac{d}{da} F^{-1}(a) = \frac{1}{f(F^{-1}(a))}
\]

Substituting back, the derivative of user surplus simplifies to:

\[
(\Delta u - F^{-1}(a))
\]

Now, differentiate the full welfare function:

\[
\frac{dW}{da} = (\Delta u - F^{-1}(a)) + \frac{d}{da}(v(1 - a)) + \frac{d}{da}(am)
\]

\[
\frac{dW}{da} = (\Delta u - F^{-1}(a)) - v + m
\]

Setting the first-order condition to zero:

\[
\Delta u - F^{-1}(a^*) - v + m = 0 \quad \Rightarrow \quad F^{-1}(a^*) = \Delta u - v + m
\]

\begin{equation}
a^* = F(\Delta u - v + m)
\label{eq:6}
\end{equation}

Compare to the private equilibrium $a^{\text{priv}} = F(\Delta u)$. Since $F$ is strictly increasing (as $f(c) > 0$):

\[
a^{\text{priv}} > a^* \iff \Delta u > \Delta u - v + m \iff v - m > 0 \iff v > m
\]

Thus, whenever $v > m$, the private delegation exceeds the socially optimal level, producing excessive delegation and a corresponding deadweight loss.

\textit{Part 3 (Information Asymmetry Cost)}  
The information asymmetry forces a single price \( p(a) \) for impressions of heterogeneous quality. This creates a systematic mispricing where the price reflects the \textit{average} quality of traffic, not the \textit{individual} quality of an impression. This distortion has two sides:

The first side is human impressions, which are high-quality. For a human-generated impression, which advertisers value at \( v \), they pay the market price \( p(a) = v(1 - a) \). Because the price is below the true value, advertisers realize a surplus on this impression, calculated as:

\[
\text{Value} - \text{Price} = v - p(a) = v - v(1 - a) = va
\]

The second side is AI agent-generated impressions, which are low-quality. For an AI-generated impression, which advertisers value at 0, they still pay the market price \( p(a) = v(1 - a) \). Because the price is above the true value, advertisers realize a loss on this impression, calculated as:

\[
\text{Value} - \text{Price} = 0 - p(a) = -v(1 - a)
\]

While the pricing for any single impression is distorted, the market structure ensures that the advertiser's \textit{expected} surplus across all transactions is zero. The surplus gained from human impressions is exactly offset by the loss incurred from AI impressions:

\[
\mathbb{E}[\text{Surplus}] = (1 - a)(va) + a(-v(1 - a)) = va - va^2 - va + va^2 = 0
\]

While advertisers break even on average, the market is inefficient. The cost of information asymmetry is not a net loss to advertisers but a distortion to the pricing mechanism. High-quality assets (human attention) are priced too low, and low-quality assets (AI attention) are priced too high. This is a classic lemons market problem, which as shown in Part 2—leads to allocative inefficiency and deadweight loss. Ultimately, the cost is borne by publishers through lost revenue, and by society through misallocated attention and productivity. \quad$\blacksquare$

\section{Publisher Adaptation Strategies}

Faced with eroding ad revenues from the \textit{attention lemons} problem, publishers must consider strategic adaptations to survive. To analyze their core decision, I first define the competitive environment. As established in the model, publishers operate as monopolistic competitors; each offers differentiated content, granting them a degree of market power. This differentiation becomes particularly salient when considering autonomous AI agents, which are often task-driven and seek specific, non-substitutable information.
For such agents, a publisher's content is not easily replaced. or such agents, a publisher’s content is not easily replaced.
For instance, an AI agent tasked with retrieving the methodology from a specific scientific paper
published in Nature cannot fulfill its query by visiting social media or a video platform. It requires the unique, authoritative content from the original source. This specificity makes the agent's demand for access to that particular publisher highly inelastic. Consequently, I formalize this insights with an assumption about the nature of this market power.

\textbf{Assumption A:}\label{assump:market_power}
\textit{In the short run, a true and single monopolistic publisher $i$'s choice of toll, $\tau_i$, is assumed to leave both the market-wide AI share, $a$, and the publisher's own total traffic share, $s_i$, invariant. Formally}

\[
\frac{\partial a}{\partial \tau_i} = 0 \quad \text{and} \quad \frac{\partial s_i}{\partial \tau_i} = 0.
\]

\hyperref[assump:market_power]{Assumption A} posits that a unilateral change in a single publisher’s AI toll leaves both the market-wide delegation share and that publisher’s traffic share locally unchanged. The first part, $\frac{\partial a}{\partial \tau_i} = 0$, is a standard application of monopolistic competition theory, as established by \textcite{dixit_stiglitz1977monopolistic}, where an individual firm is atomistic and its unilateral actions have a negligible first-order effect on market-wide aggregates like the overall AI delegation rate. The second condition, $\frac{\partial s_i}{\partial \tau_i} = 0$, is a short-run simplification, invoked for analytical tractability, that reflects the fact that, for task-driven AI agents, the publisher’s unique content lacks a perfect substitute. Faced with technical switching costs, a rational AI-agent developer will prefer to pay toll rather than allow the service to fail, which renders demand for that specific publisher’s content perfectly inelastic. More formally, consider a rational AI developer who needs to access content from publisher $i$ to fulfill a user query. The value generated by successfully fulfilling this query using publisher $i$'s differentiated content is
\[
B.
\]

When publisher $i$ imposes a per-query toll of $\tau_i$, the developer must choose between paying the toll or refusing.

If the developer pays the toll, its net payoff for the query is
\[
V_{\text{pay}} = B - \tau_i.
\]

If the developer refuses to pay, it cannot access publisher $i$'s content and must resort to its best outside option. In the short run, these options are:

\begin{enumerate}[label=\roman*.]
    \item Redirecting the query to an imperfect alternative source, which yields a lower value $B_{\text{alt}} < B$, or
    \item Failing to fulfill the query, which incurs a direct cost $C_F > 0$ from reputational damage and potential user churn.
\end{enumerate}

A rational developer will choose the better of these two failure outcomes. Thus, the payoff from the outside option is
\[
V_{\text{out}} = \max(B_{\text{alt}}, -C_F).
\]

The developer will choose to pay the toll if and only if the payoff from paying exceeds the payoff from its outside option, i.e.,
\[
V_{\text{pay}} > V_{\text{out}}.
\]

This gives the condition:
\[
B - \tau_i > \max(B_{\text{alt}}, -C_F).
\]

This inequality can be rearranged to find the developer's maximum willingness to pay, which can be defined as the threshold toll $\tau_{\text{max}}$:
\[
\tau_i < B - \max(B_{\text{alt}}, -C_F) \equiv \tau_{\text{max}}.
\]

The term $\tau_{\text{max}}$ represents the value premium of publisher $i$'s specific content over the developer's next best alternative. For any toll $\tau_i$ set by the publisher such that
\[
0 \leq \tau_i < \tau_{\text{max}},
\]
the developer will always choose to pay. Consequently, the quantity of AI traffic to publisher $i$,
\[
s_i^{AI},
\]
remains constant within this range. This establishes a region of perfectly inelastic demand for the publisher's content access. If the toll were to equal or exceed $\tau_{\text{max}}$, demand would drop to zero as the developer switches to its outside option.

Therefore, the assumption
\[
\frac{\partial s_i}{\partial \tau_i} = 0
\]

holds, if the publisher's profit-maximizing toll, $\tau_i^*$, lies strictly within this inelastic region. As derived in  \hyperlink{prop:tolling_optimality}{Proposition 2} , the publisher's optimal toll under the Tolling strategy is given by
\[
\tau_i^* = \theta_i a s_i.
\]

Therefore, the assumption is valid if and only if
\[
\tau_i^* < \tau_{\text{max}}.
\]

Substituting the expressions for these terms yields the formal condition:
\[
\theta_i a s_i < B - \max(B_{\text{alt}}, -C_F).
\]

While this condition enables analytical tractability, it also limits the generality of the model. For example, a publisher with exceptionally high tolling efficiency ($\theta_i$) or operating in a market with widespread AI adoption ($a$) may optimally choose a toll that exceeds the inelasticity threshold, thereby violating the assumption. Similarly, in a long-run setting where developers can invest in reducing switching costs or cultivating perfect and better alternatives, the threshold $\tau_{\text{max}}$ may diminish over time. Nonetheless, the assumption remains a reasonable approximation within the short-run context of the model.

Now, under this assumption, a rational publisher may choose one of three strategies to maximize revenue. I assume a publisher may only choose one. In the null action (N) strategy, the publisher continues to monetize its total traffic share $s_i$ at the diluted market price $p(a)$. The revenue is unchanged from the initial model:

\[
R^N_i = s_i p(a) = s_i v (1 - a)
\]

In the blocking (B) strategy, the publisher implements technology to exclude all AI agents from its site at a fixed cost $k_i > 0$. An important feature of the ``attention lemons'' market is that an individual publisher cannot unilaterally change the market-wide price $p(a)$. Even if a publisher successfully creates a ``human-only'' stream of traffic, they must still sell those impressions into the general market, which cannot perfectly verify this quality and thus continues to offer the blended price $p(a)$.

This action alters the publisher's traffic volume. The publisher loses all traffic that was generated by AI agents ($a s_i$). However, I introduce a user retention parameter, $\lambda \in [0,1]$, representing the fraction of users who had delegated browsing but choose to revert to visiting the site manually once the agent is blocked. The publisher's new total traffic volume is its original human traffic, $(1 - a)s_i$, plus the retained traffic, $\lambda a s_i$. The total traffic is therefore $s_i (1 - a + \lambda a)$, or $s_i (1 - a(1 - \lambda))$.

The revenue function for the Blocking strategy:

\begin{equation}
R^B_i = v \cdot s_i (1 - a(1 - \lambda))(1 - a) - k_i
\label{eq:7}
\end{equation}

In the tolled API (T) strategy, the publisher permits access by AI agents but imposes a toll $\tau > 0$ per AI query. The publisher continues to earn the diluted ad price $p(a)$ on its total traffic $s_i$ and additionally collects toll revenue from its share of AI traffic, $a s_i$. The publisher incurs a convex implementation cost $c_i(\tau)$. As derived in \hyperlink{prop:tolling_optimality}{Proposition 2}, setting the optimal toll $\tau^*_i$ yields the following maximal revenue:

\begin{equation}
R^T_i(\tau^*_i) = s_i v (1 - a) + \frac{1}{2} \theta_i a^2 s_i^2
\label{eq:8}
\end{equation}

where $\theta_i > 0$ is a parameter measuring the publisher’s tolling efficiency.
\medskip

\noindent
\textbf{PROPOSITION 1 (Dominance of the Tolling Strategy)}  \\
\textit{For any share of AI traffic $a > 0$, the Tolling (T) strategy strictly dominates both the Null (N) and the Blocking (B) strategies. It is the unique rational strategy for a publisher.}

\medskip
\noindent\textbf{Proof.}
The proof proceeds in two parts. First, I show that the Null strategy is strictly dominated by Tolling. Second, I show that the Blocking strategy is also strictly dominated by Tolling.\\

\textit{Part 1: Tolling Dominates Null Action}\\
A publisher prefers Tolling to Null if $R^T_i > R^N_i$:

\[
s_i v(1 - a) + \frac{1}{2} \theta_i a^2 s_i^2 > s_i v(1 - a)
\]

Subtracting $s_i v(1 - a)$ from both sides yields:

\[
\frac{1}{2} \theta_i a^2 s_i^2 > 0
\]

By assumption, $\theta_i > 0$, $s_i > 0$, and $a > 0$. The inequality is therefore always true. The Null strategy is strictly dominated.

\medskip
\textit{Part 2: Tolling Dominates Blocking}\\
A publisher prefers Tolling to Blocking if $R^T_i > R^B_i$. We substitute the formula for $R^B_i$:

\[
s_i v(1 - a) + \frac{1}{2} \theta_i a^2 s_i^2 > s_i (1 - a(1 - \lambda)) v(1 - a) - k_i
\]

We can expand the first term on the right-hand side:

\[
s_i v(1 - a) + \frac{1}{2} \theta_i a^2 s_i^2 > \left[s_i v(1 - a) - s_i a(1 - \lambda) v(1 - a)\right] - k_i
\]

The term $s_i v(1 - a)$ appears on both sides and can be cancelled. This leaves:

\[
\frac{1}{2} \theta_i a^2 s_i^2 > -s_i a(1 - \lambda) v(1 - a) - k_i
\]

The left-hand side (LHS) is strictly positive, as shown in Part 1.

The right-hand side (RHS) is the sum of two terms. The first term, $-s_i a(1 - \lambda) v(1 - a)$, is non-positive because all its constituent parts ($s_i$, $a$, $v$, $(1 - a)$, $(1 - \lambda)$) are non-negative. The second term, $-k_i$, is strictly negative by assumption ($k_i > 0$). The sum of a non-positive and a strictly negative term is strictly negative. The inequality requires that a strictly positive number (LHS) be greater than a strictly negative number (RHS). This is always true. Therefore, the Tolling strategy strictly dominates the Blocking strategy.   \quad$\blacksquare$

Given that the Tolling strategy is the unique rational choice for a publisher as per Proposition 1, I now solve for the profit-maximizing toll. \\

\noindent
\textbf{\hypertarget{prop:tolling_optimality}{PROPOSITION 2 (Optimal Tolling)}} \\
\textit{If a publisher adopts a tolling strategy, it sets an optimal toll $\tau_i^*$ to maximize $R_i^T(\tau)$. Suppose the tolling cost function is quadratic:}

\[
c_i(\tau) = \frac{\tau^2}{2\theta_i}
\]

\textit{where $\theta_i > 0$ measures tolling efficiency. The publisher’s optimal toll is:}

\begin{equation}
\tau_i^* = \theta_i a s_i
\label{eq:10}
\end{equation}

\textit{The resulting optimal revenue from tolling is:}

\[
R_i^T(\tau_i^*) = s_i v (1 - a) + \frac{1}{2} \theta_i a^2 s_i^2
\]

\vspace{1em}

\textbf{Proof.}

The publisher chooses $\tau$ to solve:

\[
\max_{\tau} \left[ s_i v (1 - a) + a s_i \tau - \frac{\tau^2}{2 \theta_i} \right]
\]

The first-order condition with respect to $\tau$ is:

\[
\frac{\partial R_i^T}{\partial \tau} = a s_i - \frac{\tau}{\theta_i} = 0
\]

Solving for $\tau$ yields the optimal toll:

\[
\tau_i^* = \theta_i a s_i
\]

Substituting $\tau_i^*$ back into the revenue function:

\[
R_i^T(\tau_i^*) = s_i v (1 - a) + a s_i (\theta_i a s_i) - \frac{(\theta_i a s_i)^2}{2 \theta_i}
\]

\[
= s_i v (1 - a) + \theta_i a^2 s_i^2 - \frac{\theta_i^2 a^2 s_i^2}{2 \theta_i}
\]

\[
= s_i v (1 - a) + \frac{1}{2} \theta_i a^2 s_i^2
\]

This completes the proof. \quad$\blacksquare$

\section{Externality Internalization}

The negative externality imposed by AI traffic on publisher revenue arises because the private decision to delegate browsing does not account for the social cost of that action, specifically the resulting degradation of the advertising market. I now show that this market failure can be corrected and the socially efficient level of delegation, $a^*$, can be sustained in a private equilibrium through a mechanism that forces users to internalize this externality.

\bigskip

\noindent
\textbf{THEOREM 2 (Pigouvian Correction of the Delegation Externality)}\\
\textit{If a per-delegation cost $\tau$ is imposed on each user who delegates browsing to an AI agent, the socially efficient level of delegation $a^*$ can be sustained as a private equilibrium by setting the cost to
\[
\tau^* = v - m.
\]
This corrective mechanism is viable whenever $v > m$, the same condition under which inefficiently excessive delegation occurs.}

\bigskip

\noindent\textbf{Proof.} The proof proceeds by modifying the private delegation decision and solving for the cost $\tau$ that aligns the resulting private equilibrium with the social optimum derived in Section~\ref{sec:lemons_failure}.

First, I analyze the choice of a single human user. In the presence of a per-delegation cost $\tau \geq 0$, a user with a private cost of delegation $c$ chooses to delegate to an AI agent if and only if the net utility from doing so remains positive. The user's decision rule is:
\[
\Delta u - c - \tau > 0 \quad \Leftrightarrow \quad c < \Delta u - \tau.
\]

The aggregate share of users who choose to delegate, which determines the share of AI traffic $a$, is therefore a function of the cost $\tau$:
\[
a(\tau) = \mathbb{P}(c < \Delta u - \tau) = F(\Delta u - \tau),
\]
where $F(\cdot)$ is the cumulative distribution function of the private cost $c$.

Next, recall from the analysis of social welfare in Section~\ref{sec:lemons_failure} that the socially optimal level of delegation $a^*$, which maximizes total welfare, is given by:
\[
a^* = F(\Delta u - v + m).
\]

To restore efficiency, the cost $\tau$ must be set such that the private equilibrium level of delegation $a(\tau)$ equals the socially optimal level $a^*$. I set $a(\tau) = a^*$ and solve for the optimal cost $\tau^*$:
\[
F(\Delta u - \tau^*) = F(\Delta u - v + m).
\]

By assumption, the density function $f(c)$ is strictly positive, which implies that the cumulative distribution function $F(\cdot)$ is strictly increasing. Therefore, the equality holds if and only if the arguments of the function $F$ are equal:
\[
\Delta u - \tau^* = \Delta u - v + m.
\]

Solving for $\tau^*$ yields:
\[
\tau^* = v - m.
\]

This is the optimal corrective cost. It forces the marginal user to internalize the net social cost of their delegation decision. The social cost is the advertiser value $v$ lost from that impression, and the social benefit is the AI provider’s profit $m$. The difference, $v - m$, is the precise magnitude of the marginal negative externality.

This cost $\tau^*$ can be implemented in two ways. It can be viewed as a Pigouvian tax levied by a central authority on each act of delegation, with the revenue potentially used to compensate publishers. Alternatively, it can be interpreted as the outcome of a Coasean bargain. If publishers have the right to block AI agents, they can collectively demand a per-query access fee
\[
\tau_{\text{agent}} = v - m
\]
from AI agents. A profit-maximizing AI agent would, in turn, pass this cost on to its users, setting a user-facing fee
\[
\tau_{\text{user}} = \tau_{\text{agent}},
\]
thus implementing the optimal cost $\tau^*$ and restoring market efficiency. \quad$\blacksquare$

\section{Tipping Point and Market Collapse}

The preceding analysis establishes that a continuous erosion of advertising value occurs as the share of AI traffic, $a$, rises. However, the dynamics of digital ecosystems often exhibit nonlinear collapse beyond critical thresholds. I now formalize the tipping point at which the ad-funded content economy ceases to function.\noindent
As established in Proposition 1, the Tolling strategy strictly dominates all other options. A rational publisher's revenue is therefore always determined by the return from Tolling:''

\begin{equation}
R^*_i(a) = R^T_i(a) = s_i v(1 - a) + \frac{1}{2} \theta_i a^2 s_i^2
\end{equation}

Each publisher faces a fixed operating cost $F_i > 0$ and remains active only if its revenue covers this cost. The advertising market collapses when the set of active publishers becomes empty. 

\bigskip
\noindent
\textbf{PROPOSITION 3 (Critical Threshold)} \\
\textit{There exists a critical delegation share $a_c \in (0,1]$ such that for all $a > a_c$, no publisher finds it profitable to operate and the advertising market collapses. This threshold $a_c$ is implicitly defined by the condition that the most resilient publisher just covers its operating cost:}
\[
\max_i \left\{ R_i^*(a_c) - F_i \right\} = 0.
\]

A critical threshold $a_c$ represents a stable tipping point only if an increase in AI traffic beyond this point causes the marginal publisher’s revenue to fall, ensuring market exit. Formally, for the marginal publisher $j$ where $R_j^*(a_c) = F_j$, it must be that $R_j^*(a)$ is locally non-increasing at $a_c$. This provides a necessary stability condition,

\[
\frac{\partial R_j^*}{\partial a_c} \leq 0,
\]

which I use to analyze the properties of the market.

\medskip
\noindent\textbf{Proof.} To prove the existence of a critical threshold $a_c$, we first define a function $\Pi(a)$ that represents the maximum profit earned by any publisher $i \in \{1, \dots, N\}$ for a given share of AI traffic $a \in [0, 1]$. As established in Proposition 1, a rational publisher’s only strategy is Tolling. Therefore, the maximum profit function is given by:

\[
\Pi(a) = \max_i \left\{ R_i^T(a) - F_i \right\}
\]

The market remains viable as long as $\Pi(a) \geq 0$, as this implies at least one publisher is earning non-negative profits. The market collapses when $\Pi(a) < 0$. The critical threshold $a_c$ is therefore a root of the equation $\Pi(a) = 0$. We establish the existence of such a root using the Intermediate Value Theorem.

First, we must establish that $\Pi(a)$ is a continuous function on the compact interval $[0, 1]$.

A publisher’s optimal revenue function $R_i^*(a)$ is the revenue from Tolling:

\[
R_i^*(a) = R_i^T(a) = s_i v (1 - a) + \frac{1}{2} \theta_i a^2 s_i^2
\]

This revenue function is a quadratic polynomial in $a$ and is therefore continuous for all $a$. The profit function for publisher $i$, $R_i^*(a) - F_i$, is the difference between a continuous function and a constant, and is thus also continuous. The function $\Pi(a)$ is the maximum of a finite set of $N$ continuous functions, which guarantees that $\Pi(a)$ is itself continuous on $[0, 1]$.

Second, evaluate $\Pi(a)$ at the boundaries of the interval.

At $a = 0$, there is no AI traffic. The publisher’s revenue is:

\[
R_i^*(0) = s_i v (1 - 0) + 0 = v s_i
\]

The maximum profit in the market is:

\[
\Pi(0) = \max_i \{v s_i - F_i\}
\]

For the ad-funded market to be economically viable in its initial state, I make the non-triviality assumption that at least one publisher is profitable in the absence of AI traffic. Thus, assume:

\[
\Pi(0) > 0
\]

At $a = 1$, all traffic is generated by AI agents. The publisher’s revenue is determined entirely by what it can earn from tolls:

\[
R_i^*(1) = s_i v (1 - 1) + \frac{1}{2} \theta_i (1)^2 s_i^2 = \frac{1}{2} \theta_i s_i^2
\]

The maximum profit in the market is:

\[
\Pi(1) = \max_i \left\{ \frac{1}{2} \theta_i s_i^2 - F_i \right\}
\]

We assume that fixed costs $F_i$ are sufficiently large such that no publisher can remain profitable when earning revenue only from AI-generated traffic via tolls. This economically sensible condition ensures that:

\[
\Pi(1) < 0
\]

We have established that $\Pi(a)$ is a continuous function on $[0, 1]$, and under reasonable economic assumptions, $\Pi(0) > 0$ and $\Pi(1) < 0$. By the Intermediate Value Theorem, there must exist at least one $a_c \in (0, 1)$ such that $\Pi(a_c) = 0$.

Let $a_c$ be the smallest such root. By this construction:

\begin{enumerate}[label=\roman*.]
    \item For all $a \in [0, a_c)$, $\Pi(a) > 0$ and the market is viable.
    \item At $a = a_c$, the most resilient publisher breaks even.
    \item For any $a > a_c$, $\Pi(a) < 0$, implying no publisher is profitable and the market has collapsed.
\end{enumerate}

This proves the existence of the critical threshold. \quad$\blacksquare$
\bigskip

\noindent
\textbf{COROLLARY 1 (Stability Properties)} \\
\textit{The critical threshold for market collapse, $a_c$, is determined by the economic parameters of the dominant Tolling strategy. Its stability properties are as follows:}

\begin{enumerate}[label=\roman*.]
    \item The threshold $a_c$ increases with the advertiser valuation for human attention ($v$) and the tolling efficiency ($\theta_j$) of the marginal publisher.
    \item The threshold $a_c$ decreases with the marginal publisher's fixed operating costs ($F_j$).
\end{enumerate}

\noindent \textbf{Proof.}  Let the marginal publisher at the threshold be indexed by $j$. This publisher is defined as the most resilient firm, whose profit is zero at the critical threshold $a_c$. From Proposition 1, the Tolling strategy is the unique dominant strategy, so the profit function for publisher $j$ is

\[
\Pi_j(a) = R_j^T(a) - F_j,
\]

where $R_j^T(a)$ is the revenue from tolling and $F_j$ is the fixed operating cost. The threshold $a_c$ is therefore the root of the implicit function $G(a_c, \boldsymbol{x}) = 0$, where $\boldsymbol{x}$ is the vector of parameters $(v, \theta_j, F_j)$:

\[
G(a_c, \boldsymbol{x}) = s_j v (1 - a_c) + \frac{1}{2} \theta_j a_c^2 s_j^2 - F_j = 0.
\]

To derive the comparative statics using the Implicit Function Theorem, we must first establish the sign of $\frac{\partial G}{\partial a_c}$.
The profit function $\Pi_j(a)$ is a quadratic function of $a$, which can be written as:

\[
\Pi_j(a) = \left( \frac{1}{2} \theta_j s_j^2 \right) a^2 - (s_j v) a + (s_j v - F_j).
\]

By assumption, $\theta_j > 0$ and $s_j > 0$, so the coefficient on the $a^2$ term is strictly positive. This establishes that $\Pi_j(a)$ is a strictly convex and continuous function on $a \in [0, 1]$.

The existence of a market collapse threshold in Proposition 3 is predicated on two economic assumptions:

\begin{enumerate}[label=\roman*.]
    \item The publisher is profitable in the absence of AI traffic, $\Pi_j(0) = s_j v - F_j > 0$;
    \item The publisher is unprofitable when all traffic is from AI agents,
    \[
    \Pi_j(1) = s_j v (0) + \frac{1}{2} \theta_j s_j^2 - F_j < 0.
    \]
\end{enumerate}

Given that $\Pi_j(a)$ is a continuous function on $[0, 1]$ with $\Pi_j(0) > 0$ and $\Pi_j(1) < 0$, the Intermediate Value Theorem guarantees the existence of at least one root $a_c \in (0, 1)$. The strict convexity of $\Pi_j(a)$ ensures this root is unique. For a strictly convex function to pass from a positive value to a negative value over an interval, its slope at the unique root within that interval must be strictly negative. Thus, it is a necessary property of the model that

\[
\left. \frac{\partial \Pi_j}{\partial a} \right|_{a = a_c} < 0.
\]

This implies that the denominator in the Implicit Function Theorem, $\frac{\partial G}{\partial a_c}$, is strictly negative.

The remaining partial derivatives of $G$ are:

\begin{enumerate}[label=\roman*.]
    \item $\displaystyle \frac{\partial G}{\partial v} = s_j (1 - a_c)$. As $a_c \in (0, 1)$, this derivative is strictly positive.
    \item $\displaystyle \frac{\partial G}{\partial \theta_j} = \frac{1}{2} a_c^2 s_j^2$. As $a_c > 0$, this derivative is strictly positive.
    \item $\displaystyle \frac{\partial G}{\partial F_j} = -1$.
\end{enumerate}

Applying the Implicit Function Theorem, $\displaystyle \frac{\partial a_c}{\partial x_k} = - \frac{\partial G / \partial x_k}{\partial G / \partial a_c}$,  find:

\begin{enumerate}[label=\roman*.]
    \item $\displaystyle \frac{\partial a_c}{\partial v} = \frac{- s_j (1 - a_c)}{- s_j v + \theta_j a_c s_j^2} = \frac{(-)}{(-)} > 0.$
    \item $\displaystyle \frac{\partial a_c}{\partial \theta_j} = \frac{- \frac{1}{2} a_c^2 s_j^2}{- s_j v + \theta_j a_c s_j^2} = \frac{(-)}{(-)} > 0.$
    \item $\displaystyle \frac{\partial a_c}{\partial F_j} = \frac{-1}{- s_j v + \theta_j a_c s_j^2} = \frac{(-)}{(-)} < 0.$
\end{enumerate}

An increase in advertiser valuation $v$ or tolling efficiency $\theta_j$ increases market resilience by raising the threshold $a_c$. An increase in fixed costs $F_j$ decreases market resilience by lowering $a_c$. The parameters of the dominated Blocking strategy do not affect the market’s tipping point. \quad$\blacksquare$

\section{Conclusion}
This paper develops a theoretical framework to analyze the economic consequences of the rise of autonomous AI agents in two-sided advertising markets. By formalizing the concept of an \textit{attention lemons} problem, I demonstrate how the substitution of human browsing with non-monetizable AI agent traffic creates a fundamental market failure. The analysis reveals that as the share of AI-generated traffic grows, the expected value of an ad impression declines, leading to a direct and continuous erosion of publisher revenue.

The resulting single price market systematically misprices attention assets: human engagement is undervalued while synthetic interactions are overvalued, generating deadweight loss. For profit-maximizing publishers, both the simple laissez-faire (null action) strategy and the blocking strategy are strictly dominated by the tolling strategy.

To address the root externality, the paper proposes a Pigouvian correction. Whether implemented as a tax or a market-driven fee structure, this intervention ensures that users bear the true cost of their delegation decisions, restoring the market to a socially efficient equilibrium. In the absence of such a systemic correction, the model shows that the market's resilience is finite. A critical threshold exists beyond which publisher revenues no longer cover operating costs, precipitating a collapse of content supply and market function.

The analysis is primarily static and assumes homogeneity among agents and advertisers. Relaxing the assumption of homogeneous AI agents would introduce valuable theoretical insights, for instance by allowing for a subset with purchasing intent (e.g., shopping bots). This would alter the structure of the price function \(p(a)\) and show how agent heterogeneity could mitigate the mispricing distortion and reshape the optimal Pigouvian correction. 

Furthermore, the paper relies on an assumption regarding the short-run inelasticity of demand for individual publisher content. Relaxing this assumption to allow for elastic responses even in the short run might introduce compelling insights. In a scenario where AI agent developers face low switching costs and readily available perfect substitutes for publisher content, publishers would lose their monopolistic leverage, and the tolling strategy might no longer be strictly dominant. In the short run, this could lead to competitive pricing pressure and significantly lower tolls, potentially driving publisher revenues downward as agents migrate to alternative providers. In the long run, with even greater flexibility and the entry of substitute content providers, the market structure could evolve toward commodification, potentially eliminating publisher power entirely. However, the most promising extension lies in empirical validation. I leave these extensions for future work.

\clearpage

\printbibliography

\end{document}